\documentclass[12pt,a4paper,dvips]{article}
\usepackage{epsfig,wrapfig,times,mathptm} 
\renewcommand{\section}[1]{\vspace{6pt} \noindent\mbox{#1} \newline \noindent}
\renewcommand{\subsection}[1]{\vspace{6pt} \noindent\mbox{\underline{#1}} 
\newline \noindent}
\renewcommand{\subsubsection}[1]{\vspace{6pt} \noindent\mbox{\underline{#1}}
\noindent}

\newfont{\sansb}{cmssbx10}
\newfont{\sans}{cmss10}

\begin{document}
{\small HE 1.2.35 \vspace{-24pt}\\}     
{\center \LARGE SUPERLUMINAL PARTICLES AND HIGH-ENERGY COSMIC RAYS
\vspace{6pt}\\}
L. Gonzalez-Mestres$^{1,2}
$ \vspace{6pt}\\
{\it $^1$Laboratoire de Physique Corpusculaire, Coll\`ege de France, 
75231 Paris Cedex 05, France\\
$^2$Laboratoire d'Annecy-le-Vieux de Physique des Particules, 74941 
Annecy-le-Vieux Cedex, 
France
\vspace{-12pt}\\}
{\center ABSTRACT\\}
Lorentz symmetry has been tested at low energy with great accuracy, but
its extrapolation to very high-energy phenomena is much less well established.
We expect a possible breaking of Lorentz symmetry to be a very high
energy and very short distance phenomenon, compatible with existing data.
If textbook special relativity is only an approximate property of the
equations describing a sector of matter above some critical distance scale,
superluminal sectors of matter may exist related to new degrees of freedom
not yet discovered experimentally. The new superluminal particles
({\bf "superbradyons"}) would 
have positive mass and energy, and behave kinematically
like "ordinary" particles (those with critical speed in vacuum equal to 
$c$ , the
speed of light) apart from the difference in critical speed 
(we expect $c_i~\gg ~c$ , where $c_i$ is the critical
speed of a superluminal sector
of matter).
At speed $v~ >~ c$ ,
they are expected to release "Cherenkov" radiation ("ordinary" particles) in
vacuum. If superluminal particles exist, they could provide most of the cosmic
(dark) matter and produce very high-energy cosmic rays compatible with
unexplained discoveries reported in the literature. We discuss: a) the possible
relevance of superluminal matter to the composition, sources and spectra of
high-energy cosmic rays; b) signatures and experiments allowing to possibly
explore such effects.

\setlength{\parindent}{1cm}
\section{RELATIVITY, MATTER AND SUPERLUMINAL PARTICLES}
Lorentz invariance can be viewed as a symmetry of the motion
equations, in which case no reference to absolute
properties of space and time is required and the
properties of matter play the main role
(Gonzalez-Mestres, 1996). In a two-dimensional
galilean space-time,
the equation:
\equation
\alpha ~\partial ^2\phi /\partial t^2~-~\partial ^2\phi /\partial x^2 = F(\phi )
\endequation
with $\alpha$ = $1/c_o^2$ and $c_o$ = critical
speed, remains unchanged under "Lorentz" transformations leaving
invariant the squared
interval:
\equation
ds^2 = dx^2 - c_o^2 dt^2
\endequation
so that matter made with solutions of equation (1)
would feel a relativistic space-time even if the real space-time is actually
galilean and if an absolute rest frame exists in the
underlying dynamics beyond the wave equation.
A well-known example is provided by the solitons of the sine-Gordon equation,
obtained taking in (1):
\equation
F(\phi ) = (\omega /c_o)^2~sin~\phi
\endequation
where $\omega $ is a characteristic frequency of the dynamical system.
A two-dimensional universe made of sine-Gordon solitons plunged
in a galilean world would behave like a two-dimensional minkowskian
world with the laws of special relativity.
Information on any absolute rest frame would be lost by the solitons, as
if the Poincar\'e relativity principle (Poincar\'e, 1905)
were indeed a law of Nature, even if
actually the
basic equation derives from a galilean world with an
absolute rest frame. The actual structure of space and time
can only be found by going beyond the wave equation
to deeper levels of resolution, similar to the way 
high-energy accelerator experiments explore the inner structure of
"elementary" particles. At this stage, a  crucial question arises
(Gonzalez-Mestres, 1995):
is $c$ (the speed of light) the only critical speed in vacuum, are
there particles with a critical speed different from that of light?
The question clearly makes sense, as in a
perfectly transparent crystal it is possible to identify
at least two critical speeds: the speed of light and
the speed of sound. It has been shown (Gonzalez-Mestres,
1995 and 1996) that superluminal
sectors of matter can be consistently
generated, with the conservative choice of leaving the Planck constant
unchanged, replacing in the Klein-Gordon equation the
speed of light by a new critical speed $c_i$ $\gg $ $c$
(the subscript $i$ stands for the $i$-th superluminal sector). All
standard kinematical concepts and
formulas (Schweber, 1961) remain correct, leading to particles with
positive mass and energy which are not tachyons. 
We shall call them {\bf superbradyons} as, according to standard 
vocabulary (Recami, 1978), they are bradyons with superluminal critical speed in vacuum. The rest energy
of a superluminal particle of mass $m$ and critical speed $c_i$
will be given by the generalized Einstein equation:
\equation
E_{rest}~~=~~m~c_i^2
\endequation
Energy and momentum conservation will in principle not be
spoiled by the existence of several critical speeds in vacuum:
conservation laws will as usual hold for phenomena leaving the vacuum 
unchanged. Each superluminal sector will have its own Lorentz invariance
with $c_i$ defining the metric.
Interactions between two different
sectors will break both Lorentz invariances. Lorentz
invariance
for all sectors simultaneously will at best be explicit 
(i.e. exhibiting the diagonal sectorial Lorentz metric) in a single
inertial frame ({\bf the vacuum rest frame}, i.e. the "absolute" rest
frame). In our approach, the Michelson-Morley result is not incompatible
with the existence of some "ether" as
suggested by recent results in particle physics: if
the vacuum is a material medium where fields
and order parameters can condense, it may well
have a local rest frame. If superluminal particles couple
weakly to ordinary matter, their effect on the ordinary sector will occur at
very high energy and short distance (Gonzalez-Mestres, 1997a), far from
the domain of successful
conventional tests of Lorentz invariance 
(Lamoreaux, Jacobs, Heckel, Raab and Forston,
1986 ;
Hills and Hall, 1990). In particular, superbradyons naturally escape the
constraints on the critical speed derived in some specific models
(Coleman and Glashow, 1997; Glashow, Halprin, Krastev, Leung and 
Pantaleone, 1997).
High-energy experiments can therefore
open new windows in this field.
Finding some track of a superluminal sector (e.g. through
violations of Lorentz invariance in the ordinary sector) may
be the only way to experimentally discover the vacuum rest frame.
Superluminal particles lead to consistent cosmological models 
(Gonzalez-Mestres, 1997b), where they may well provide
most of the cosmic (dark) matter. Although recent criticism 
to this suggestion has been emitted in a specific model
on the grounds of gravitation theory (Konstantinov, 1997),  
it should be noticed that the framework used is 
crucially different
from the multi-graviton approach suggested in our papers.

\section{IMPLICATIONS FOR HIGH-ENERGY COSMIC RAYS}
The kinematical properties and Lorentz transformations
of high-energy superluminal particles have been
discussed in a previous paper (Gonzalez-Mestres, 1997c).
If an absolute rest frame exists, Lorentz contraction is a real physical
phenomenon and is governed by the factor 
$\gamma _i^{-1}~=~(1~-~v^2c_i^{-2})^{1/2}$ 
for the $i$-th superluminal sector,
so that there is no Lorentz singularity when a superluminal particle crosses
the speed value $v~=~c$ . Similarly, if superbradyons have any coupling to the
electromagnetic field, we expect the magnetic force to be proportional to
$v~c_i^{-1}$ instead of $v~c^{-1}$ .
Contrary to tachyons, superbradyons
can emit "Cherenkov" radiation (i.e. particles with lower 
critical speed) in vacuum. 
If $c_i~\gg ~c$ , and if the vacuum rest frame is close to that defined  
requiring isotropy of cosmic microwave background radiation,
high-energy superluminal particles will be seen on earth 
as traveling mainly at speed $v~\approx ~10^3~c$ .  
A superluminal particle moving with velocity ${\vec {\mathbf v}}$ with
respect to the vacuum rest frame,
and emitted by an astrophysical object, can reach an observer
moving with laboratory
speed  ${\vec {\mathbf V}}$ with respect to the same frame, at a time (as
measured by the observer) previous to the emission time. Such a phenomenon will
happen if  ${\vec {\mathbf v}}.{\vec {\mathbf V}}~>~c^2$ , and the
emitted particle will be seen to evolve backward in time (but it evolves
forward in time in the vacuum rest frame, so that the reversal of the time 
arrow is not really a physical phenomenon). 
If they interact several times with
the detector,
superluminal particles can
be a directional probe preceding the detailed observation of
astrophysical phenomena, such as explosions releasing
simultaneously neutrinos, photons and superluminal particles.
For a high-speed
superluminal cosmic ray with critical speed $c_i~\gg ~c$ ,
the momentum, as measured in the laboratory,
does not provide directional
information on the source, but on the vacuum rest frame.
Velocity provides directional information on the source,
but can be measured only if the
particle interacts several times with the detector, which is far from
guaranteed, or if the superluminal particle is associated to a collective
phenomenon emitting also photons or neutrinos simultaneously.
In the most favourable case,
directional detection of high-speed
superluminal particles in a large underground or
underwater detector would allow
to trigger a dedicated astrophysical observation in the direction of the sky
determined by the velocity of the superluminal particle(s).
If $d$ is the distance between the observer and the astrophysical object,
and $\Delta t$ the time delay between the detection of the superluminal
particle(s) and that of photons and neutrinos,
we have: $d~\simeq ~c~\Delta t$ .

Annihilation of pairs of superluminal particles into ordinary ones can
release very large kinetic energies and provide a new
source of high-energy cosmic rays. 
Decays of superluminal particles may play a similar role.
Collisions (especially, inelastic with very large energy
transfer) of high-energy superluminal particles with
extra-terrestrial ordinary matter may also yield high-energy
ordinary cosmic rays. Pairs of slow superluminal particles can also
annihilate into particles of another superluminal sector
with lower $c_i$ , converting most of the rest energies into a
large amount of kinetic energy.
Superluminal particles moving at $v~>~c$ can release anywhere "Cherenkov"
radiation in vacuum, i.e.
spontaneous emission of particles of a
lower critical speed $c_i$ (for $v~>~c_i$) including ordinary ones,
providing a new source of (superluminal or ordinary)
high-energy cosmic rays.
High-energy superluminal particles
can directly reach the earth and undergo
collisions inside the atmosphere, producing many secondaries like
ordinary cosmic rays. They can also interact with the rock or
with water near some underground or underwater detector,
coming from the atmosphere or after having crossed the earth,
and producing clear signatures.
Contrary to neutrinos, whose flux is strongly attenuated by the
earth at
energies above $10^6$ $GeV$ , superluminal particles will in
principle not be stopped by earth at these energies.
In inelastic collisions, high-energy superluminal primaries can
transfer most of their energy to ordinary particles.
Even with a very weak interaction probability,
and assuming that the superluminal primary does not produce any ionization,
the rate for superluminal cosmic ray events can be observable
if we are surrounded by important concentrations of superluminal
matter. Background rejection would be further enhanced by
atypical ionization properties.

The possibility that superluminal matter exists, and that it plays
nowadays an important role in our Universe,
should be kept in mind when addressing the two basic questions
raised by the analysis of any cosmic ray event:
a) the nature and properties of the cosmic ray primary; b)
the identification (nature and position) of the source of the cosmic ray.
If the primary
is a superluminal particle, it will escape conventional criteria
for particle identification
and most likely produce a specific signature
(e.g. in inelastic collisions) different from those of ordinary
primaries.
Like neutrino events, in the absence of
ionization (which will in any case be very weak) we
may expect the event to start anywhere inside the detector.
Unlike very high-energy neutrino events,
events created by superluminal primaries can
originate from a particle having crossed the earth.
An incoming, relativistic superluminal particle with momentum $p$ and
energy $E_{in} \simeq p~ c_i $ in the vacuum 
rest frame, hitting an ordinary particle at rest,
can, for instance, release most of its energy into
two ordinary particles with
momenta (in the vacuum rest frame)
close to $p_{max}~=~1/2~p~c_i~c^{-1}$ and oriented back
to back in such a way
that the two momenta almost cancel. 
Then, an energy $E_R \simeq E_{in} $
would be transferred to ordinary secondaries.
Corrections due to the earth motion
must be applied (Gonzalez-Mestres, 1997c) before defining the expected 
event configuration in laboratory experiments (AUGER, AMANDA...).
At very high energy, such events
would be easy to identify in large volume detectors, even at very small rate.
If the
source is superluminal, it can be located anywhere
(and even be a free particle) and will not necessarily be at the same place as
conventional sources
of ordinary cosmic rays. High-energy cosmic ray events
originating form superluminal sources
will provide hints on the location of such
sources and be possibly the only way to observe them.
The energy dependence of the events should be taken into account.
At very high energies,
the Greisen-Zatsepin-Kuzmin (GZK) 
cut-off (Greisen, 1966; Zatsepin and Kuzmin, 1966) 
does not in principle hold for
cosmic ray events originating from superluminal matter:
this is obvious if the primaries are superluminal particles
that we expect to interact very weakly with the cosmic microwave
background,
but is also true for ordinary primaries as we do not expect them to
be produced at the locations of ordinary sources and there is no upper
bound to their energy around $100~EeV$. 
Besides "Cherenkov" deceleration, a superluminal cosmic
background radiation may exist
and generate its own GZK
cutoffs for the superluminal sectors. However, if there are large amounts
of superluminal matter around us, they can be the main superluminal source
of cosmic rays reaching the earth.
To date, there is
no well established interpretation of the
highest-energy cosmic ray events. 
Primaries (ordinary or superluminal)
originating from superluminal particles are acceptable candidates
and can possibly escape several problems
(event configuration, source location, energy dependence...)
faced by cosmic rays produced at ordinary sources.

\section{ACKNOWLEDGEMENTS}
It is a pleasure to thank J. Gabarro-Arpa, as well as the Director of
LAPP M. Yvert and colleagues at LPC Coll\`ege de France, for useful
discussions and remarks.

\section{REFERENCES}
\setlength{\parindent}{-5mm}
\begin{list}{}{\topsep 0pt \partopsep 0pt \itemsep 0pt \leftmargin 5mm
\parsep 0pt \itemindent -5mm}
\vspace{-15pt}
\item
Coleman, S. and Glashow, S.L., "Cosmic Ray and Neutrino Tests of
Special Relativity", paper
hep-ph/9703240 of LANL (Los Alamos) electronic archive (1997).
\item
Glashow, S.L., Halprin, A., Krastev, P.I., Leung, C.N. and Pantaleone, J.,
"Comments on Neutrino Tests of Special Relativity", paper hep-ph/9703454
of LANL
electronic archive (1997).
\item
Gonzalez-Mestres, L., "Properties of a Possible Class of Particles
Able to Travel Faster than Light", Proceedings of the
Moriond Workshop on "Dark Matter in Cosmology, Clocks and Tests of
Fundamental Laws", Villars January 21 - 28 1995 ,
Ed. Fronti\`eres, Gif-sur-Yvette (1995).
\item
Gonzalez-Mestres, L., "Physical and Cosmological Implications of a Possible
Class of Particles Able to Travel Faster than Light", contribution to the
28$^{th}$ International Conference on High-Energy Physics, Warsaw July 1996 .
Paper hep-ph/9610474 of LANL electronic archive (1996).
\item
Gonzalez-Mestres, L., "Lorentz Invariance and Superluminal Particles",
paper physics/9703020 of LANL electronic archive (1997a).
\item
Gonzalez-Mestres, L., "Vacuum Structure, Lorentz Symmetry and Superluminal
Particles", paper physics/9704017 of LANL
electronic archive (1997b).
\item
Gonzalez-Mestres, L., "Space, Time and Superluminal Particles",
paper physics/9702026 of LANL electronic archive (1997c).
\item
Greisen, K., {\it Phys. Rev. Lett.} 16 , 748 (1966).
\item
Hills, D. and Hall, J.L., {\it Phys. Rev. Lett.} 64 , 1697 (1990).
\item
Konstantinov, M.Yu., "Comments on the Hypothesis about Possible Class
of Particles Able to Travel faster than Light: Some Geometrical Models",
paper physics/9705019 of LANL electronic archive (1997).
\item
Lamoreaux, S.K., Jacobs, J.P., Heckel, B.R., Raab, F.J. and Forston, E.N.,
{\it Phys. Rev. Lett.} 57 , 3125 (1986).
\item
Poincar\'e, H., Speech at the St. Louis International Exposition of 1904 ,
{\it The Monist 15} , 1 (1905).
\item
Recami, E., in "Tachyons, Monopoles and Related Topics", Ed. E. Recami, 
North-Holland, Amsterdam (1978).
\item 
Schweber, S.S., "An Introduction to Relativistic Quantum Field Theory", 
Row, Peterson and Co., Evanston and Elmsford, USA (1961). 
\item
Zatsepin, G.T. and
Kuzmin, V.A., {\it Pisma Zh. Eksp. Teor. Fiz.} 4 , 114 (1966).

\end{list}

\end{document}